\documentclass[aps,prl,twocolumn,showpacs,superscriptaddress,groupedaddress]{revtex4}
\usepackage[utf8]{inputenc}
\usepackage{amsmath}
\usepackage{amsfonts}
\usepackage{graphicx}
\usepackage{xcolor}
\usepackage{epstopdf, epsfig}
\usepackage{csquotes}
\usepackage{mathtools}
\usepackage[normalem]{ulem}
\newcommand{\Bspw}{B_{\rm\scriptscriptstyle{SPW}}}
\newcommand{\hBspw}{\hat{B}_{\rm\scriptscriptstyle SPW}}

\begin{document}
\title{Ultrashort high energy electron bunches from tunable surface plasma waves driven with laser wavefront rotation}

\author{S. Marini} %\footnote{samuel.marini@polytechnique.edu}}
\affiliation{LSI, CEA/DRF/IRAMIS, \'Ecole Polytechnique, Institut Polytechnique de Paris, CNRS, F-91128 Palaiseau, France.}
\affiliation{LULI, Sorbonne Universit\'e, CNRS, \'Ecole Polytechnique, CEA, Institut Polytechnique de Paris, F-75252 Paris, France.}

\author{P. S. Kleij} %\footnote{paula.kleij@gmail.com}}
\affiliation{LSI, CEA/DRF/IRAMIS, \'Ecole Polytechnique, Institut Polytechnique de Paris, CNRS, F-91128 Palaiseau, France.}
\affiliation{LULI, Sorbonne Universit\'e, CNRS, \'Ecole Polytechnique, CEA, Institut Polytechnique de Paris, F-75252 Paris, France.}
\affiliation{Enrico Fermi Department of Physics, University of Pisa, largo Bruno Pontecorvo 3, 56127 Pisa, Italy}

\author{F. Pisani} %\footnote{f.pisani3@studenti.unipi.it}}
\affiliation{Enrico Fermi Department of Physics, University of Pisa, largo Bruno Pontecorvo 3, 56127 Pisa, Italy}

\author{F. Amiranoff}
\affiliation{LULI, Sorbonne Universit\'e, CNRS, \'Ecole Polytechnique, CEA, Institut Polytechnique de Paris, F-75252 Paris, France.}

\author{M. Grech} %\footnote{mickael.grech@polytechnique.edu}}
\affiliation{LULI, Sorbonne Universit\'e, CNRS, \'Ecole Polytechnique, CEA, Institut Polytechnique de Paris, F-75252 Paris, France.}

\author{A. Macchi} %\footnote{macchi@df.unipi.it}}
\affiliation{National Institute of Optics, National Research Council (CNR/INO), 56124 Pisa, Italy.}
\affiliation{Enrico Fermi Department of Physics, University of Pisa, largo Bruno Pontecorvo 3, 56127 Pisa, Italy}

\author{M. Raynaud} %\footnote{michele.raynaud-brun@polytechnique.edu}}
\affiliation{LSI, CEA/DRF/IRAMIS, \'Ecole Polytechnique, Institut Polytechnique de Paris, CNRS, F-91128 Palaiseau, France.}

\author{C. Riconda \footnote{caterina.riconda@upmc.fr}}
\affiliation{LULI, Sorbonne Universit\'e, CNRS, \'Ecole Polytechnique, CEA, Institut Polytechnique de Paris, F-75252 Paris, France.}

\begin{abstract}
We propose to use ultra-high intensity laser pulses with wavefront rotation (WFR) to produce short, ultra-intense surface plasma waves (SPW) on grating targets for electron acceleration. 
Combining a smart grating design with optimal WFR conditions identified through simple analytical modeling and particle-in-cell simulation allows to decrease the SPW duration (down to few optical cycles) and increase its peak amplitude. In the relativistic regime, for $I\lambda_0^2 = 3.4 \times 10^{19} {\rm W/cm^2 \mu m^2}$, such SPW are found to accelerate high-charge (few 10's of pC), high-energy (up to 70 MeV) and ultra-short (few fs) electron bunches.
\end{abstract}
\maketitle
\date{\today}

Surface plasmon polaritons, also known as surface plasma waves (SPW) in free electron media, are highly-localized electromagnetic field structures with the ability to confine and enhance light in sub-walength regions at the interface between two media \cite{raether88,barnes2003,maier2007,pitarke2007}. Their unique properties have made them ideal candidates for applications in a broad range of research fields, from bio/chemical sensing \cite{jain08,huang14} to the design of small photonic devices \cite{ozbay06,chung11}. 

The excitation of SPW by micrometric wavelength ($\lambda_0=0.8~{\rm \mu m}$) femtosecond (fs) laser pulses irradiating solid targets has been demonstrated as a strategy to enhance secondary emission of radiation and particles. 
In the low intensity regime, from few GW/cm$^2$ to tens of TW/cm$^2$, surface plasmon polaritons have lead to harmonic emission \cite{agarwal1982,coutaz1985,jatav2019} and the production of photoelectron bunches at energies up to few 100's eV \cite{kupersztych2001,zawadzka2001}.
The advent of table-top, 10's TW, fs lasers allowed on-target intensities $I_0 \lambda_0^2 \gtrsim 10^{18} {\rm W/cm^2 \mu m^2}$. In this ultra-high intensity (UHI) regime, any target material quickly turns into a plasma, and electrons reach relativistic quiver-velocities in the intense laser field. SPW then become of interest not only as unexplored nonlinear plasma modes, but also for their capability of accelerating electrons, being waves with a longitudinal electric field component and slightly subluminal phase speed. Simulations and experiments have indeed shown that relativistic SPW can accelerate high-charge, ultra-short electron bunches along the target surface \cite{cate007,ceccotti13,bigongiari13,tian,riconda,fedeli:16,fedeli2017,cantono18,macchi18,raynaud18,Zhu,raynaud19},
with energies largely exceeding their quiver-energy 
and spatio-temporal correlation with XUV harmonic emission \cite{cantono2018}.

In a recent paper, Pisani {\it et al.} \cite{pisani18} showed through electromagnetic simulations in the linear optics (low intensity) regime that using wavefront rotation (WFR) on the driving laser pulse could help generate more intense, shorter SPW. WFR is a technique used on fs lasers to induce a rotation of the successive laser wavefronts, thus leading to a time-varying incidence angle of the laser impinging onto a target. Since SPW on a grating are excited for a well defined value of this angle, using WFR allows for the SPW excitation only over a very short time, leading to the generation of near single-cycle SPW; an enhancement of the excited SPW was also found.

In this Letter, we demonstrate how these effects can be harnessed in the UHI regime, 
and  WFR can be used to drive tunable, ultrashort, ultra-intense SPW able to generate near single-cycle, highly energetic electron bunches.
The optimal WFR conditions are identified using both analytical modeling and kinetic (Particle-In-Cell, PIC) simulations.
They allow for a significant increase of both the SPW amplitude and the electron energy by up to 65\% with respect to the case without WFR. 
A careful design of the grating target allows for an additional increase (by  25\%) of the electron maximum energy.  
Electron bunches with several 10's of MeV energy and 10's of pC charge are predicted considering currently available table-top laser parameters.

\begin{figure}%[t!]
    \begin{center}
    \includegraphics[width=7.cm]{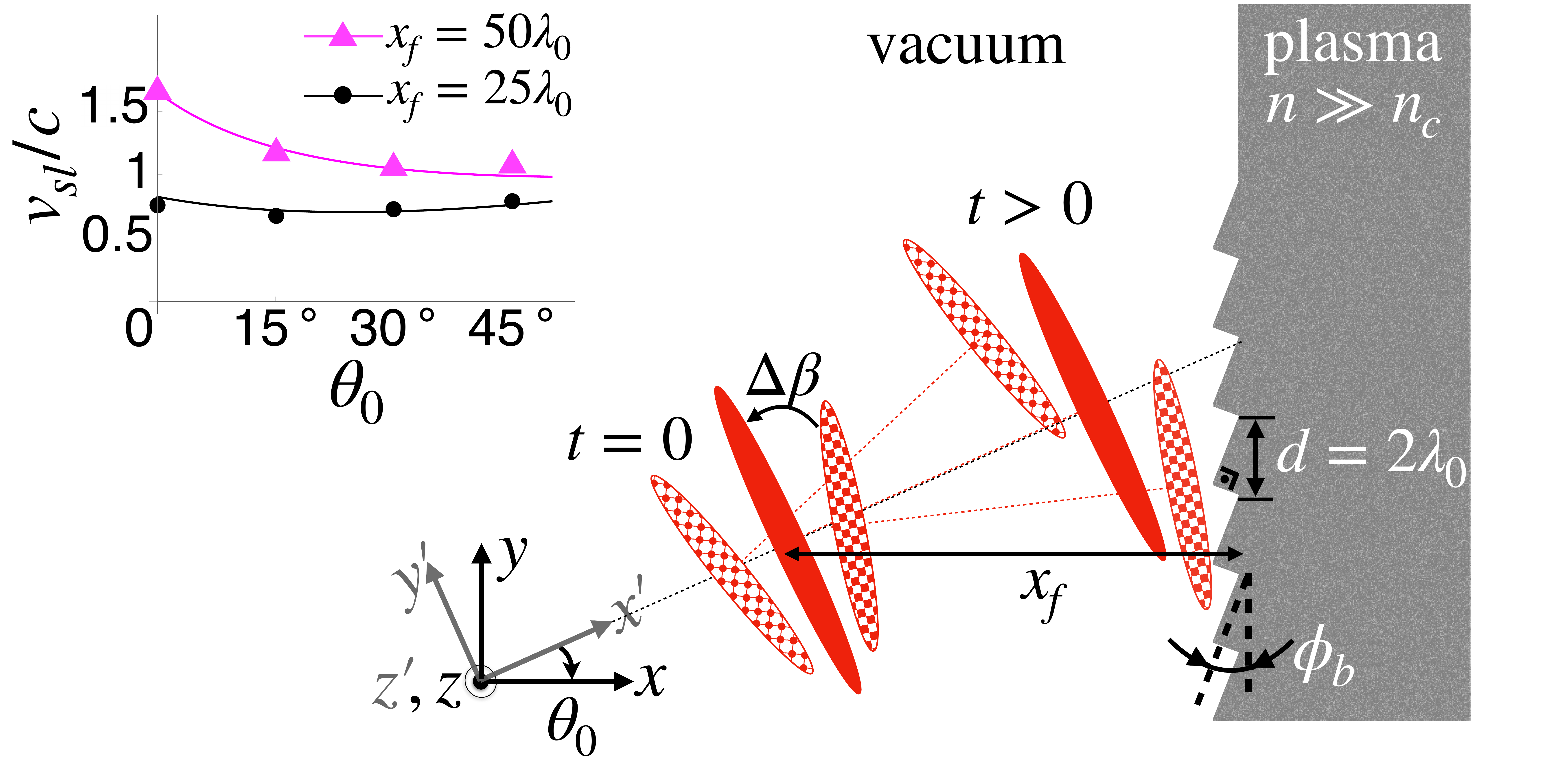}
    \caption{Interaction setup: the central laser wavefronts are shown at best focus ($t=0$), and striking the target ($t>0$). Due to WFR, setting the target at a distance $x_f$ from best focus leads to a  ``sliding focus" effect, the maximum on-target intensity sliding in the $y$-direction at a velocity $v_{\rm sl}$.
    The upper left insert compares $v_{\rm sl}$ from Eq.~\eqref{eq_vapp} (solid lines) with measures from PIC simulations (points) for $x_f=25\lambda_0$ (black) and $x_f=50\lambda_0$ (magenta).}\label{fig1} 
    \end{center}
\end{figure}

The interaction setup considered throughout this work is depicted in Fig.~\ref{fig1}. A UHI laser pulse impinges onto an overdense plasma with density $n \gg n_c$,  with $n_c=\epsilon_0m_e\omega_0^2/e^2$ the critical density at the  laser frequency $\omega_0=2\pi c/\lambda_0$, $c$ the vacuum speed of light, $\epsilon_0$ the vacuum permittivity, and $m_e$ and $-e$ the electron mass and charge. To resonantly excite a SPW at the vacuum-plasma interface, the target surface is partially modulated, and the laser incidence angle ($\theta_0$) is chosen such that $\sin \theta_0 = \sqrt{(n/n_c-1)/(n/n_c-2)} - \lambda_0/d$, with $d$ the target periodicity \citep{raether88}. The resulting SPW is excited at the laser frequency $\omega = \omega_0$, and satisfies the dispersion relation (non-relativistic cold-fluid model \cite{kaw:70}) $c^2 k{\rm\scriptscriptstyle{SPW}}^2/\omega^2=(\omega_p^2/\omega^2 - 1)/(\omega_{p}^2/\omega^2 - 2)$, with $k{\rm\scriptscriptstyle{SPW}}$ the SPW wavenumber and $\omega_{p}=\sqrt{e^2 n/(\epsilon_0 m_e)}$ the electron plasma frequency.  For $n \gg n_c$, the SPW phase and group velocities are slightly subluminal:  $v_{\phi} \rightarrow c [1-n_c/(2n)]$ and $v_g \rightarrow c [1-3n_c/(2n)]$.

As shown in Fig.~\ref{fig1}, the target is located at a distance $x_f$ from the laser best focus,
which together with WFR allows for a ``sliding focus" effect, {\it i.e.} a displacement in time of the pulse intensity peak along the target surface. If the sliding focus velocity $v_{\rm sl}$ is close to the SPW  velocity, the latter will be driven more efficiently. To estimate $v_{\rm sl}$ let us recall that  
at focus, the electric field of a pulse with WFR  can be written as \cite{vincenti12}:
\begin{equation}
\label{eq_wave}
E(y^\prime,t) = E_0\,f(t)\,F(y^\prime)\,\exp\left[i \phi(y^\prime,t) \right]\,.
\end{equation}
Here $E_0$ is the maximum electric field, $f(t)$ and $F(y^\prime)$ are the electric field temporal and transverse (in our 2D configuration) spatial envelope, and the spatio-temporal phase is
\begin{equation}
\label{eq_phase}
\phi(y^\prime,t) = \omega_0 t  \left(1 - \Omega_{\beta}\,y^\prime/c \right).
\end{equation}
The linear dependence in $y^\prime t$ leads to an instantaneous angle of propagation of light $\beta(t) \simeq -(c/\omega_0)\partial\phi/\partial y^\prime= \Omega_{\beta}t$ increasing linearly with time, with $\Omega_{\beta}$ the WFR velocity. In Fig.\ref{fig1}, $\Omega_{\beta}>0$ is considered, only the central wavefronts are represented and angles are exaggerated for illustration purposes. The main angle of incidence $\theta_0$, defined as that of the central wavefront, is chosen as the resonant angle for exciting the SPW. Successive wavefronts are then shifted by an angle $\Delta\beta = \Omega_{\beta} \lambda_0/c$ henceforth referred to as the WFR parameter. As a result, each successive wavefront will strike the target at a slightly different location along the $y$-direction leading to the apparent sliding velocity of the pulse on the target. For ultrashort pulses and/or the central wavefronts, we obtain a constant sliding velocity:
\begin{equation}
v_{\rm sl} \simeq \frac{ \Delta\beta\,x_f/\lambda_0}{\cos^2\!\theta_0 + \sin\!\theta_0 \, \Delta\beta\,x_f/\lambda_0 }\,c\,.
\label{eq_vapp}
\end{equation}
As shown in the insert of Fig.~\ref{fig1} (for $\Delta\beta=33$mrad), Eq.~\eqref{eq_vapp} is found to be in good agreement with measurements from PIC simulations \footnote{In PIC simulations, $v_{\rm sl}$ is measured by locating the position of the maximum laser field amplitude as a function of time at the target surface and  time-averaging over the laser high-frequency.}. 

The sign and value of the WFR parameter $\Delta\beta$ affects the duration and amplitude of the excited SPW \cite{pisani18}. Indeed, when the sliding velocity is along the direction of propagation of the SPW, the excited wave can increase its amplitude while maintaining a short duration. Additional tunability can be obtained by calculating an optimal value of the WFR parameter $\Delta\beta_{\rm opt}$ such that the sliding velocity $v_{\rm sl}$ coincides with the SPW velocity $\simeq c$; this leads to:

\begin{equation}
\Delta\beta_{\rm opt} \simeq \frac{\lambda_0}{x_f}\,\left( 1+\sin\theta_0 \right)\,.
\label{eq_DeltaBetaOpt}
\end{equation}
Eq.~\eqref{eq_DeltaBetaOpt} depends on $x_f$: $\Delta\beta_{\rm opt}$ decreases when increasing the distance between the target and best focus. %Equation~\eqref{eq_DeltaBetaOpt} depends on $x_f$ so that increasing the distance between the target and best focus $\Delta\beta_{\rm opt}$ is decreased.
This allows to relax the experimental constraint of obtaining large WFR velocity \cite{quere14}. However, there is a trade off since at larger values of $x_f$ the intensity of the laser at the surface decreases. For the largest value we investigate, $x_f=50\lambda_0$ [where Eq.~\eqref{eq_DeltaBetaOpt} gives $\Delta\beta_{\rm opt} \simeq 30{\rm mrad}$], the laser field amplitude on target is decreased by 8\% with respect to the configuration studied below, $x_f=25\lambda_0$ where $\Delta\beta_{\rm opt} \simeq 60{\rm mrad}$.

An additional improvement on the interaction setup was made by considering that both the efficient excitation and propagation of SPW strongly depend on the grating/surface properties. By an extensive numerical study of the effect of the target profile on the SPW excitation \cite{kleij19}, we have found that the best coupling is obtained for a blazed grating, as also suggested experimentally in \cite{cantono18}. A systematic comparison between targets fully modulated or only partially engraved showed that a partially engraved target (with grooves only in the laser-irradiated spot) efficiently mitigates radiation losses due to scattering of the SPW off the grating. The use of this mixed surface grating allows a better propagation of the SPW along the flat surface. In our simulations (not shown), we observed an increase of 25\% of the maximum electron energy using such targets.

To test our claims, two series of 2D3V PIC simulations were performed with the code {\sc Smilei}~\cite{smilei} considering different laser field strength $a_0=e E_0/(m_e c \omega_0)$.
First a non-relativistic laser intensity $a_0=0.1$ allows identify the optimal parameters for SPW excitation. Then, the UHI regime of interaction $a_0=5$ and electron acceleration along the target surface are considered .
In both cases, the general setup of the simulation is given in Fig.~\ref{fig1} with numerical parameters in~\footnote{The simulation box is $39\lambda_0 \times 72\lambda_0$ (in the $x$-$y$ directions), with $9984\times18432$ cells (spatial resolution $\Delta=\lambda_0/256$), and time resolution $\Delta t = 0.95\Delta/\sqrt{2}$. Electromagnetic field boundary conditions are injecting/absorbing in $x$ and periodic in $y$. Particle boundary conditions in $x$ are reflecting (left) or thermalizing (right), and periodic in $y$. There are $32$ macro-particles per species per cell.}. The grating target, of thickness $3\lambda_0$, has density $n=100\,n_c$, ion to electron mass ratio $m_i/(m_e) = 1836$ and temperature ratio $T_i/(T_e)= 0.1$ with  $T_e=50$eV. 
The periodicity of the grating is $d = 2\lambda_0$ with a groove’s depth $h=0.44\lambda_0$ and a blazed angle $\phi_b=13^{\circ}$. A flat surface (at $y>42\lambda_0$) follows the grating so that the laser illuminates only the number of ripples corresponding to the projected pulse waist onto the surface. The driving laser is a {\it p}-polarized Gaussian pulse with transverse size $w_\perp=5.2\lambda_0$, duration [full-width-at-half-maximum (FWHM) in intensity] $T=10\lambda_0/c$ \footnote{The laser transverse profile is Gaussian, $F(y')=\exp(-y'^2/w_\perp^2)$ with $w_\perp=5.2\lambda_0$ and its time profile is $\cos^2$: $f(t) = \cos\! \left(\pi t/(2T)\right)$ for $\vert t\vert < T$ ($0$ otherwise), with $T=10\lambda_0/c$.}. The laser pulse impinges onto the grating target at the resonant angle $\theta_0=31^{\circ}$. The simulation is run up to time  $t_0+20\lambda_0/c$, with $t_0$ the time when the peak of the pulse reaches the target. Unless specified otherwise, all values are taken at the end of the simulation. 

We first consider $a_0=0.1$ for which relativistic nonlinearities can be neglected.
The $z$-component of the magnetic field \footnote{$\Bspw$ is collected at $t=t_0 + 20\lambda_0/c$, on flat surface far from the laser-plasma interaction zone. The magnetic field has been filtered,selecting values of $k>2 k_{\rm\scriptscriptstyle{SPW}}$
.}, noted $\Bspw$ [or $\hBspw=e\Bspw/(m_e\omega_0)$], is taken as representative of the SPW, all the other field components being proportional to it.
For $n \gg n_c$, and in the vacuum side, the linear approximation yields $|E_x| \sim c|\Bspw|$ and $|E_y|\sim c|\Bspw| \sqrt{n_c/n}$.

\begin{figure}[t!]
\begin{center}
\includegraphics[width=3.3cm]{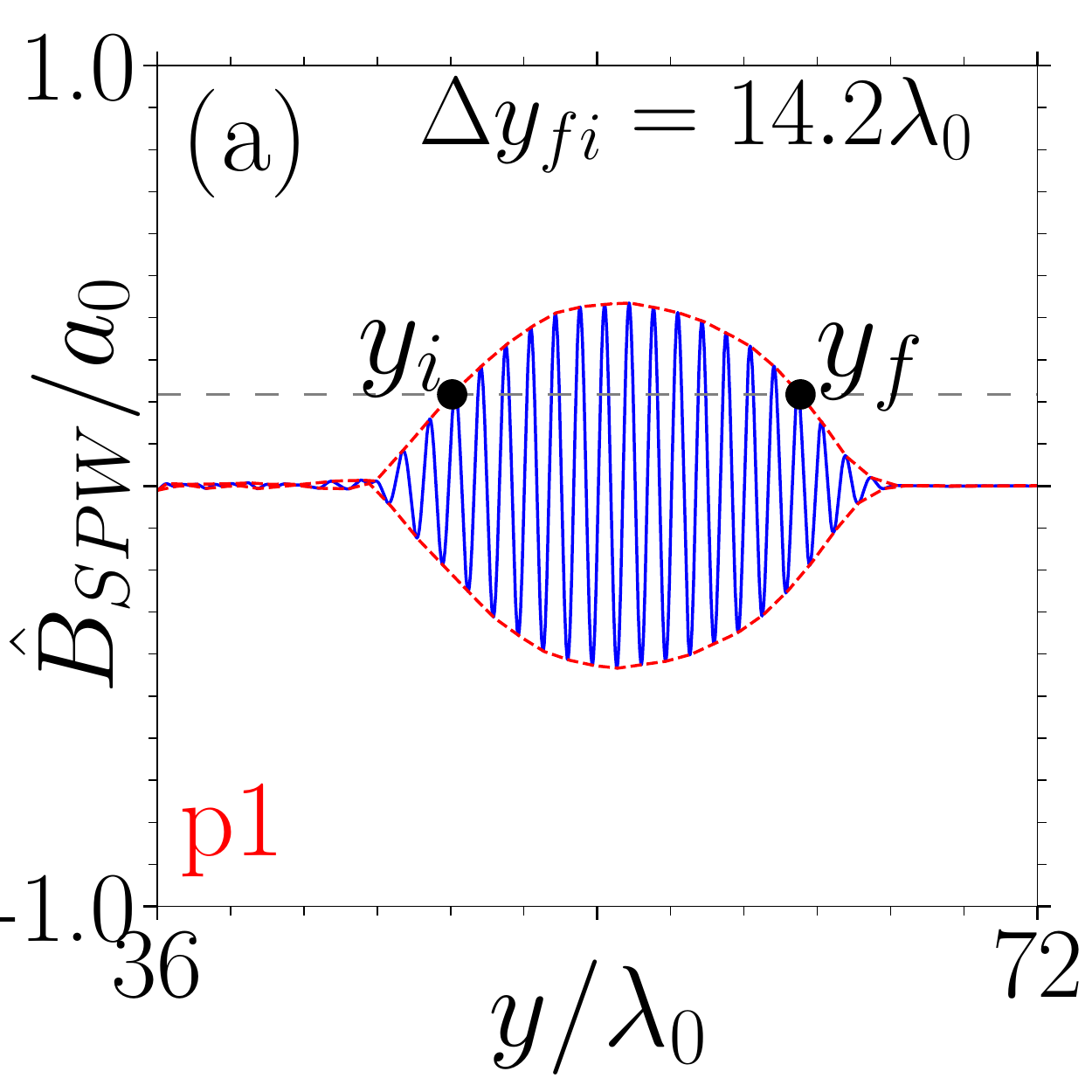}
\includegraphics[width=3.3cm]{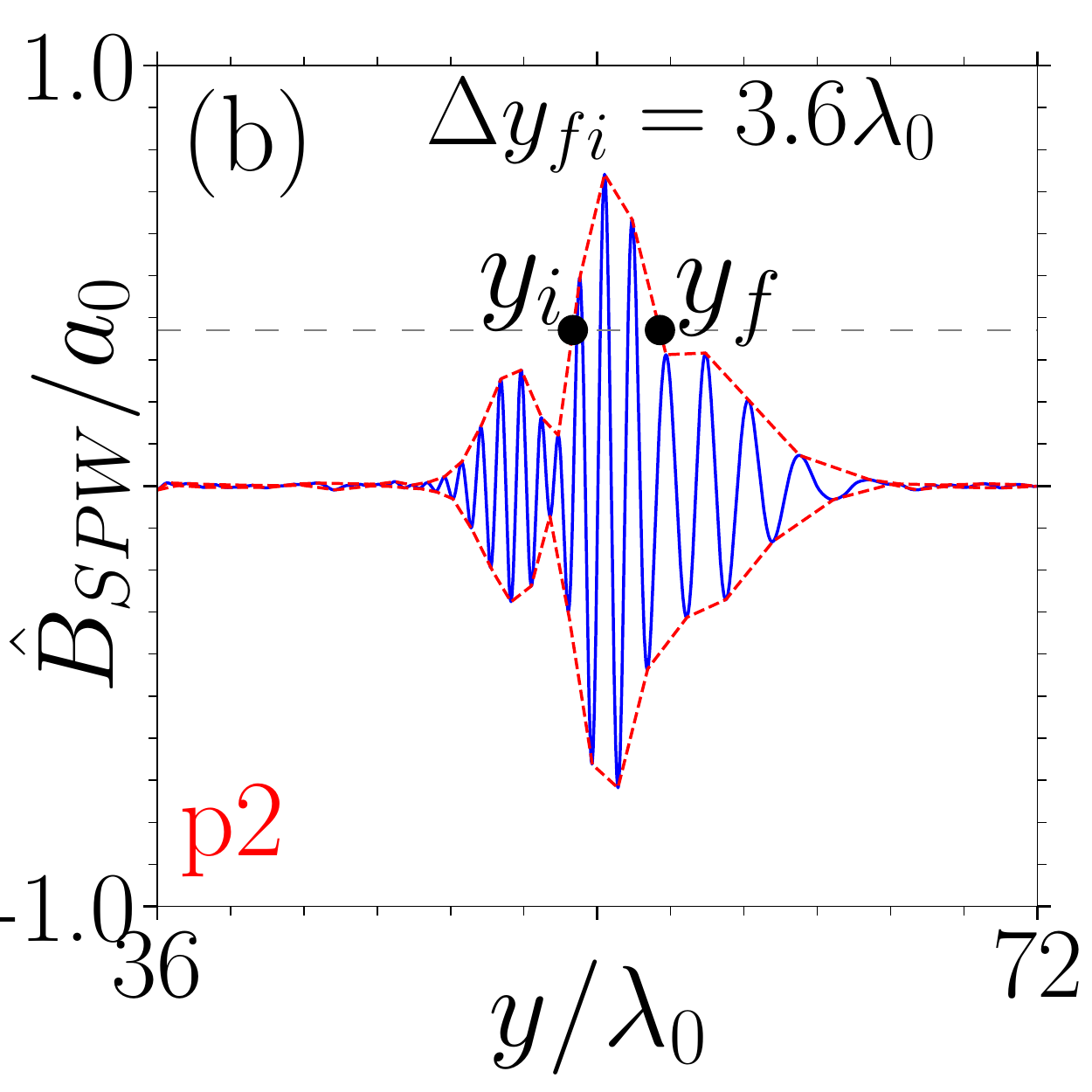}\\
\includegraphics[width=3.3cm]{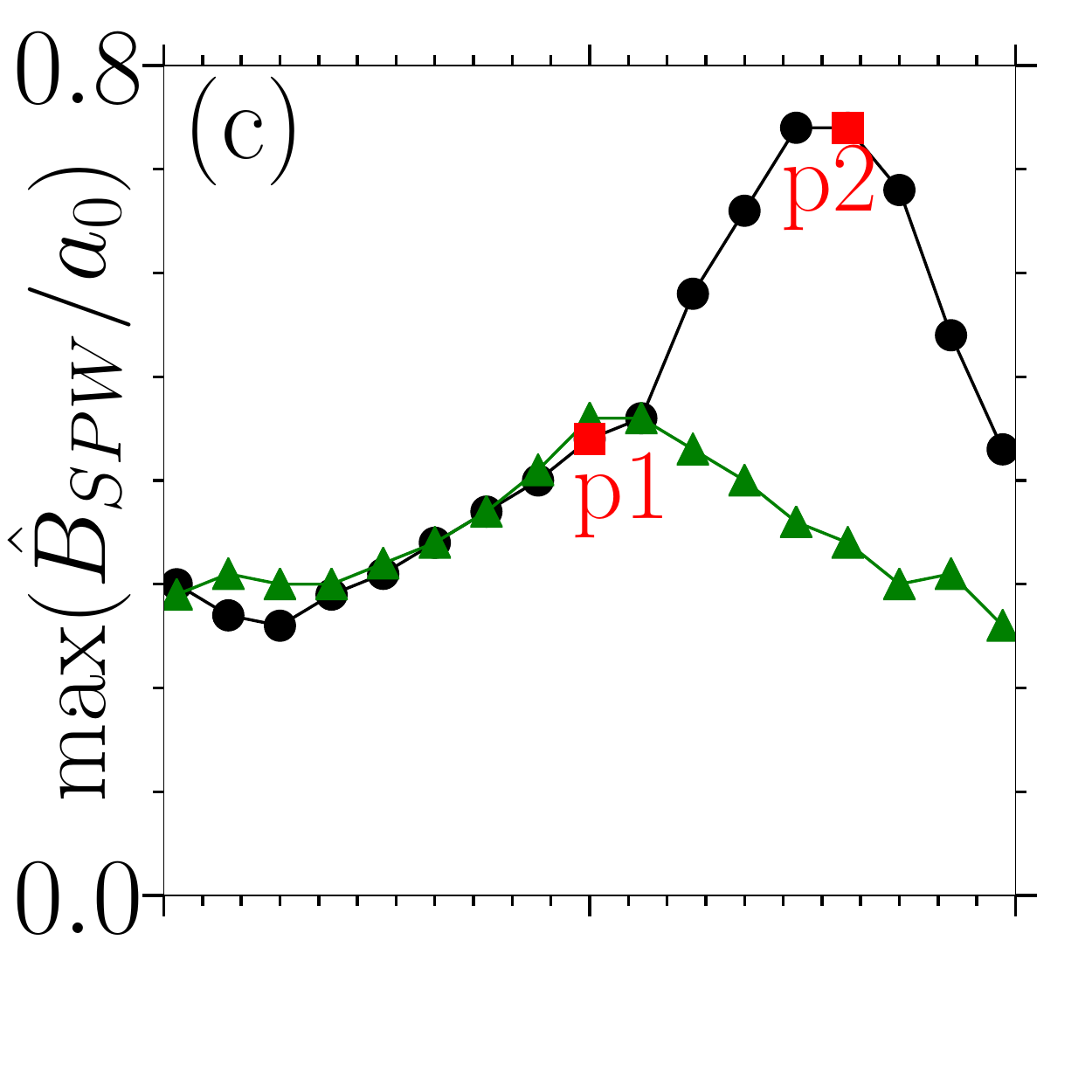}
\includegraphics[width=3.3cm]{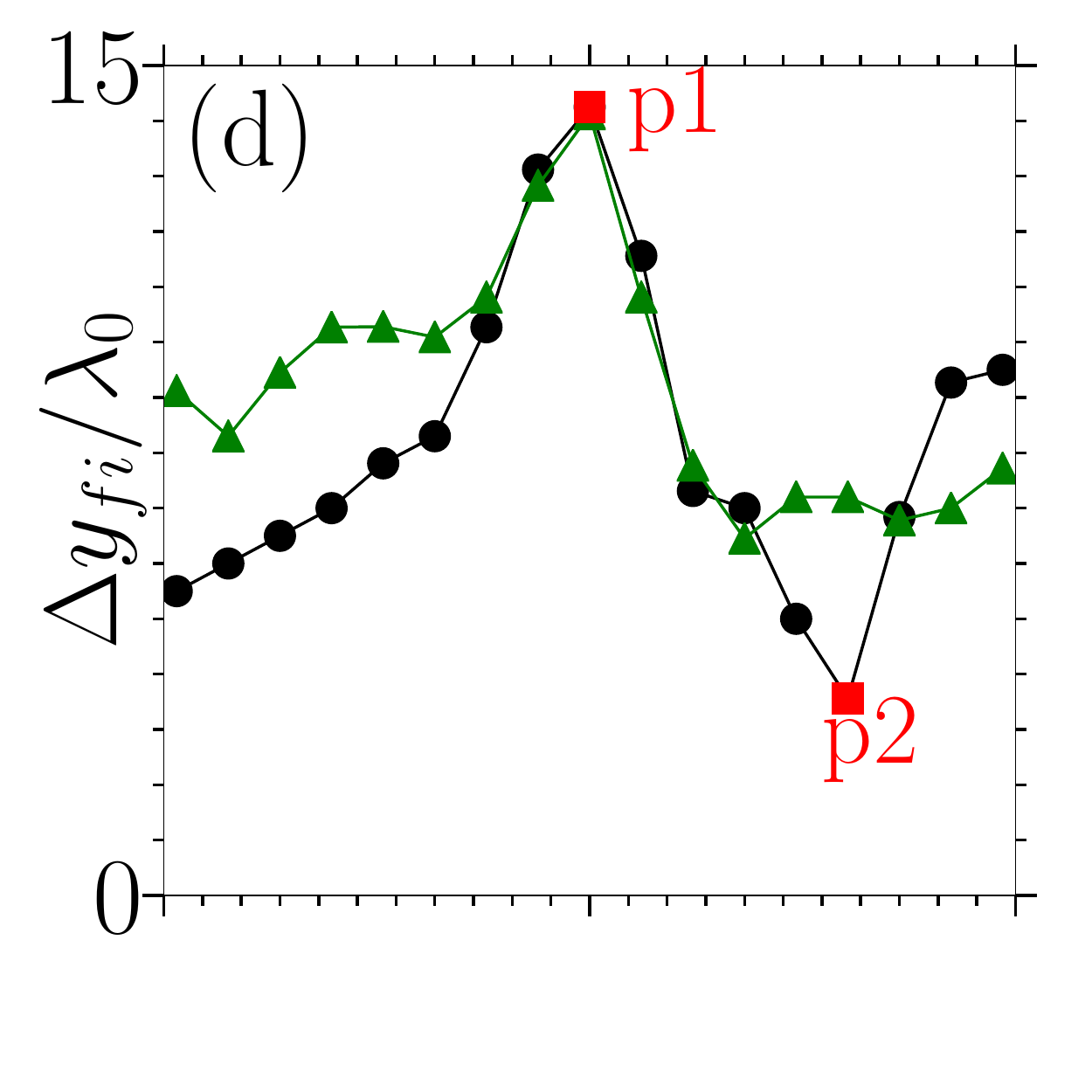}\\
\vspace{-0.45cm}
\includegraphics[width=3.3cm]{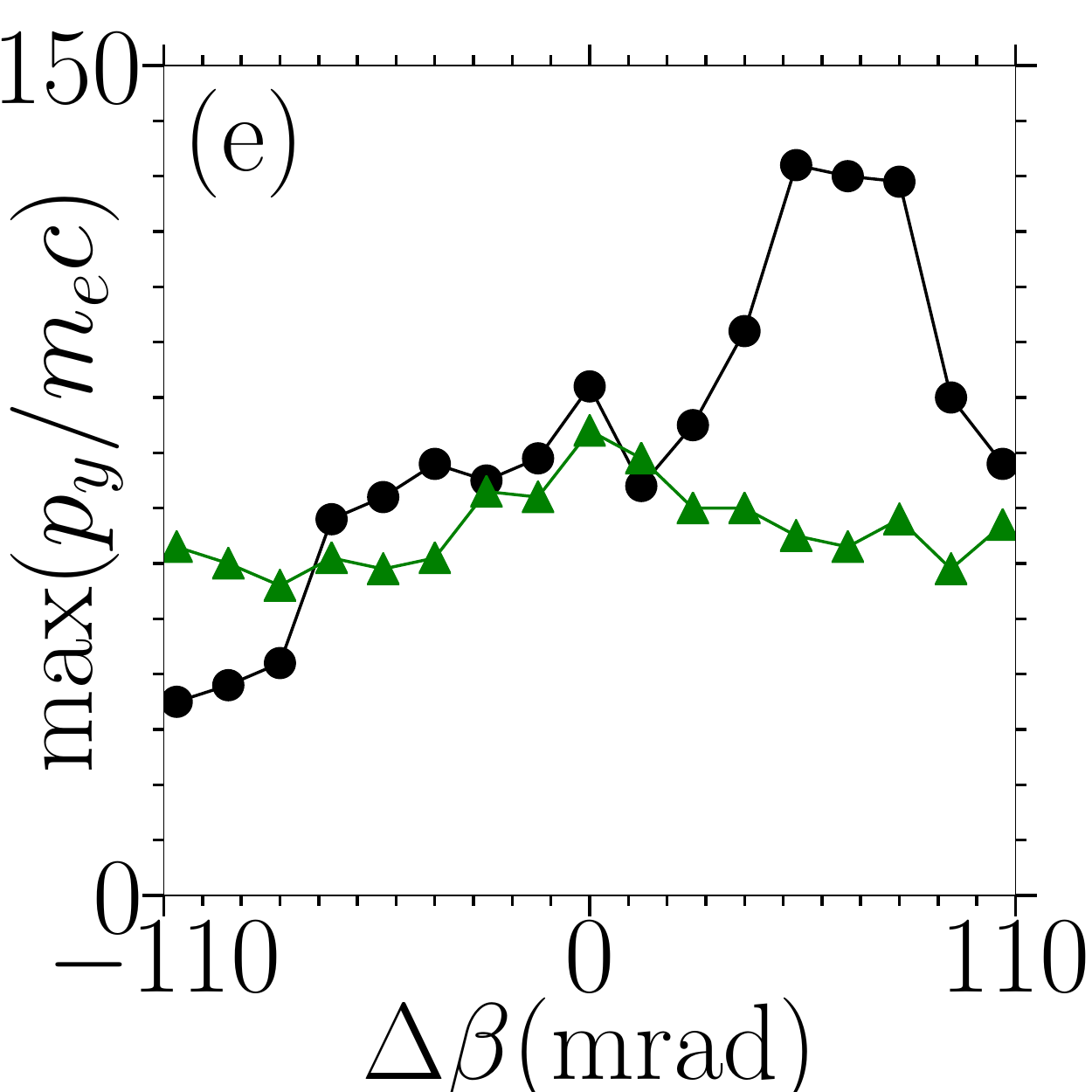}
\includegraphics[width=3.3cm]{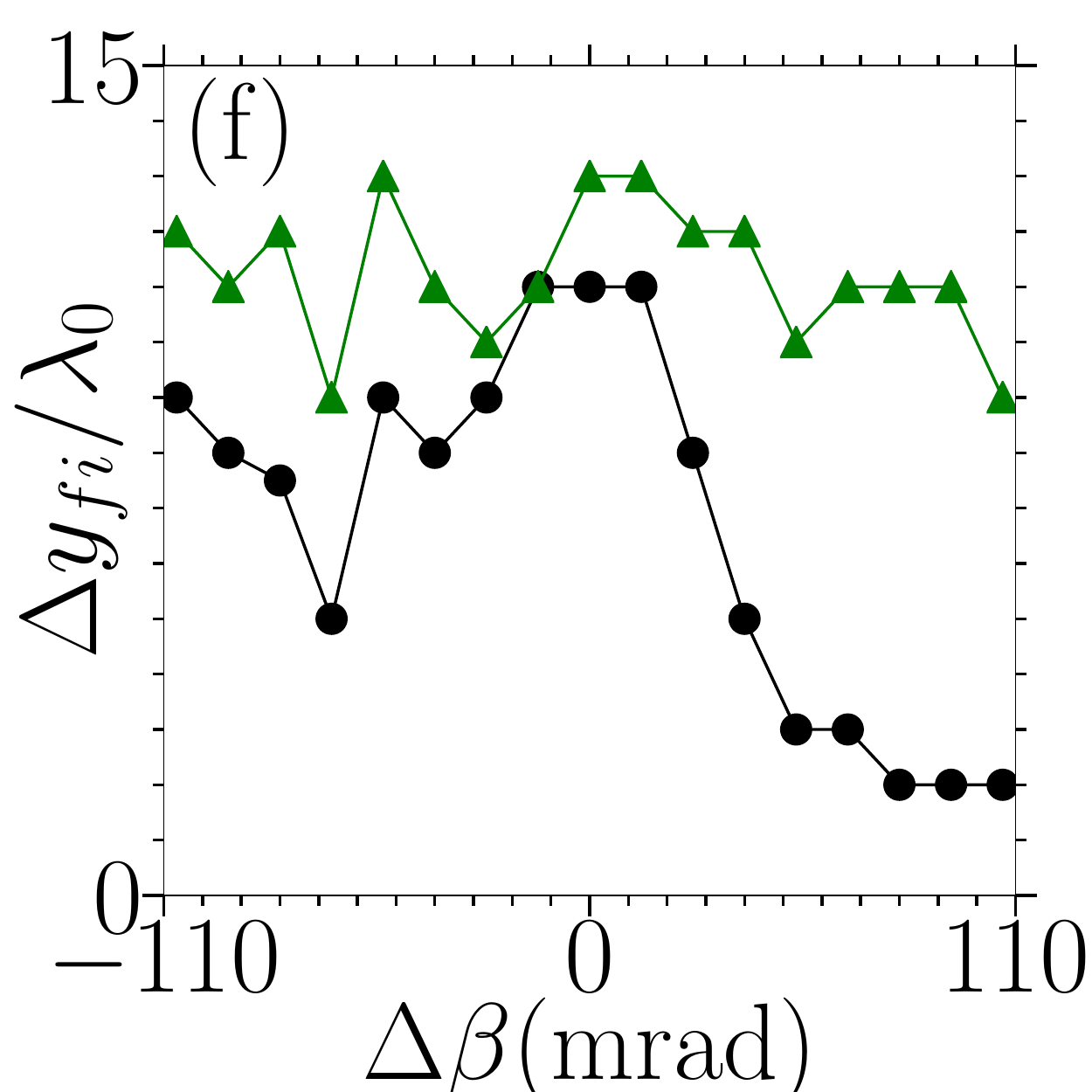}
\caption{SPW magnetic field at the target surface for (a)~$\Delta\beta=0$ and (b)~$\Delta\beta=67$mrad with $a_0 = 0.1$ and $x_f = 25\lambda_0$. (c)~Maximum SPW field amplitude and (d)~duration (FWHM) versus the WFR parameter $\Delta\beta$ for $a_0=0.1$. Markers p1 and p2 indicate the cases shown in panels~(a) and (b). (e)~Maximum electron momentum along the surface ($p_y$) and (f)~electron bunch duration (FWHM) versus $\Delta\beta$ for $a_0=5$. %At $x_f=0$ the WFR has much smaller impact on the SPW excitation and electron acceleration. 
In panels (c) to (f), $x_f=0$ (green triangles), $x_f = 25\lambda_0$ (black circles).}
\label{fig2}
\end{center}
\end{figure}

In Fig.~\ref{fig2}, we show a snapshot of $\hBspw$ along the target surface for $x_f=25\lambda_0$,
 (a)~$\Delta\beta=0$ and (b)~$\Delta\beta=67{\rm mrad}$. 
The latter case corresponds to the most intense and shortest SPW found in our simulations, 
$\Delta\beta_{\rm opt}=60{\rm mrad}$.
With this optimal WFR parameter the SPW peak amplitude is increased by $\sim 65\%$ with respect to the case without WFR and its duration, measured as the signal FWHM, is reduced by four from $14.2$ to $3.6 \lambda_0/c$.
 
Panels (c) and (d) show the maximum value of $\hBspw$ and the measured SPW duration as the result of a parametric scan of $\Delta\beta$ for $x_f=0$ (target at focus, green triangles) and $x_f=25\lambda_0$ (target off-focus, black circles). 
At focus, WFR has a small impact on the SPW excitation: 
the most intense SPW is obtained for $\Delta\beta=0$, and using non-zero $\Delta\beta$ decreases the duration of the SPW but also its maximum amplitude. 
Instead, for $x_f=25\lambda_0$, $\Delta\beta$ acts as a tuning parameter allowing both to shorten the SPW and to increase its amplitude. We observe the shortest and most intense SPW for $\Delta\beta\simeq 67{\rm mrad}$. This is in good agreement with the optimal prediction from Eq.~\eqref{eq_DeltaBetaOpt}, $\Delta\beta_{\rm opt} \simeq 60{\rm mrad}$. Note a smooth trend around this optimal value; the point directly on the left of p2 corresponds to $\Delta \beta = 53$mrad. 
Interestingly, even though the on-target laser intensity is reduced when increasing $x_f$ to $25\lambda_0$,  a significant increase of the SPW amplitude is still obtained using the optimal WFR parameter. 
A parametric scan considering $x_f=50\lambda_0$ (not shown)
leads to an optimal WFR parameter $\Delta\beta \simeq 33{\rm mrad}$ also in good agreement with
$\Delta\beta_{\rm opt}=30{\rm mrad}$ from Eq.~\eqref{eq_DeltaBetaOpt}. 
Finally, as expected positive values of $\Delta\beta$, for which the sliding velocity is along the SPW propagation direction, give a maximal effect. In contrast, for negative $\Delta\beta$, the SPW is still of a shorter duration but with a reduced amplitude, roughly that obtained when placing the target at best focus.

We now turn our attention to the second series of simulations performed in the UHI regime ($a_0=5$) and electron acceleration.
The bottom row in Fig.~\ref{fig2} shows (e)~the maximum electron momentum parallel to the surface and (f)~the characteristic width \footnote{The duration of the electron bunch is estimated from its spatial width through the relation $\Delta\tau_{fi} \simeq \Delta y_{fi}/c$.} of the accelerated electron bunch as a function of $\Delta\beta$, considering $x_f=0$ (green triangles) and $x_f=25\lambda_0$ (black circles).
Both panels exhibit very similar features than observed at low intensity.
Placing the target at focus ($x_f = 0$) the accelerated electron bunch maximum energy and duration are marginally affected by WFR.
In contrast, for $x_f=25\lambda_0$, WFR significantly impacts electron acceleration: taking $\Delta\beta > 0$ leads to more energetic, shorter electron bunches.
By comparing the case for which the target is at focus with $\Delta\beta=0$ and that with the target at $x_f=25\lambda_0$ with $\Delta\beta=67{\rm mrad}$, 
one find an increase of the maximum electron momentum by $62\%$ [from ${\rm max}(p_y) \simeq 80 m_e c$  to $\simeq 130 m_e c$] and form much shorter bunches when the optimal (positive) WFR parameter is considered and target is off-focus.  
The optimum value $\Delta\beta=67{\rm mrad}$ found for electron acceleration in this regime is the same as found earlier for efficient, ultrashort SPW excitation at lower intensity.

\begin{figure}[t!]
\begin{center}
\includegraphics[width=3.8cm]{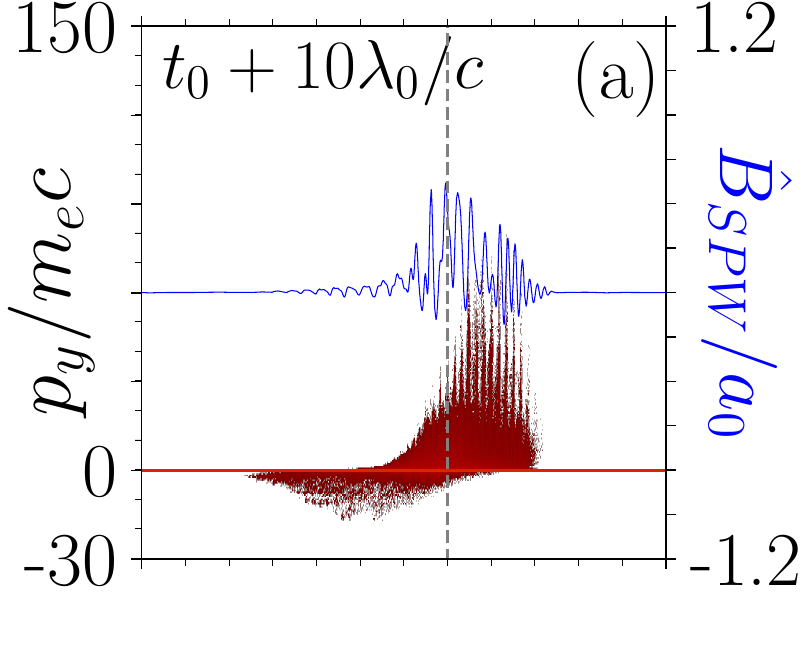}
\includegraphics[width=3.8cm]{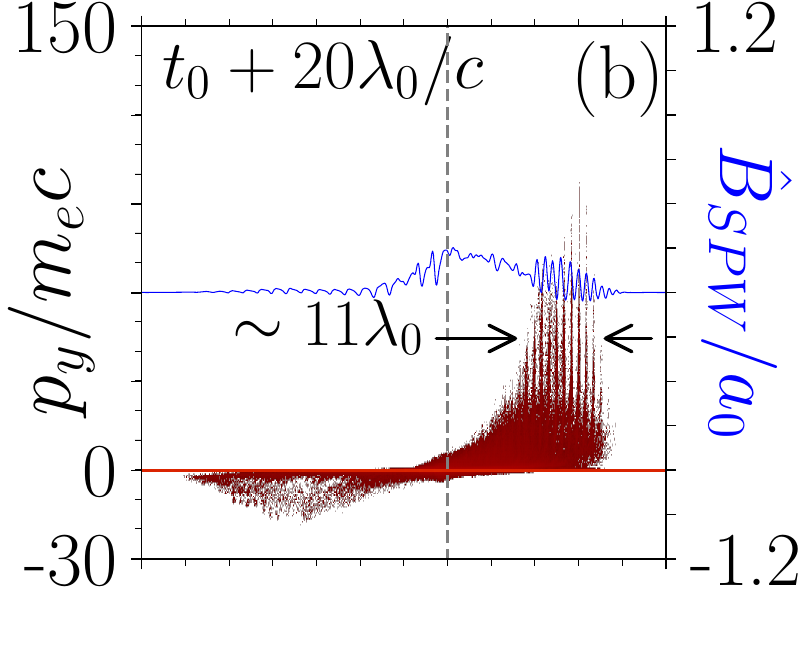}\\
\vspace{-0.27cm}
\includegraphics[width=3.8cm]{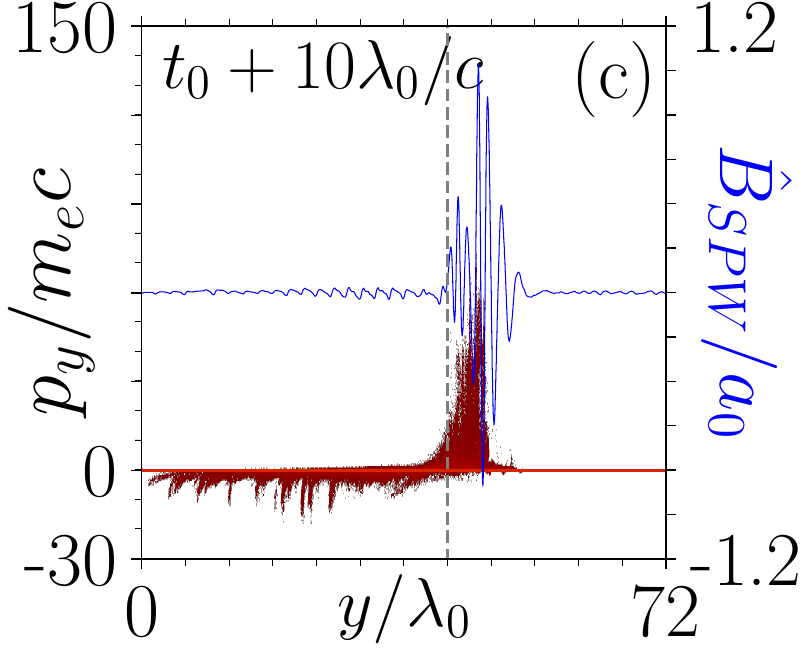}
\includegraphics[width=3.8cm]{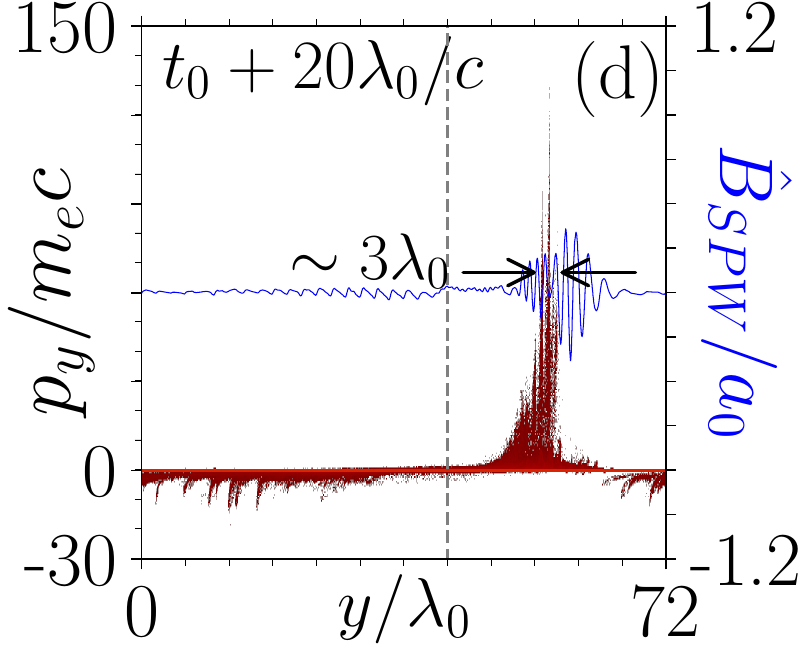}
\caption{Electron phase-space (red) and SPW field amplitude (blue line, right scale) for $a_0=5$ at times, $t=t_0+10\lambda_0/c$ and $t=t_0+20\lambda_0/c$. (a)-(b): $\Delta\beta=0$, and (c)-(d): $\Delta\beta=67{\rm mrad}$. The gray line indicates the end of the grating and beginning of the flat region.}
\label{fig3}
\end{center}
\end{figure}

Figure~\ref{fig3} gives further insights into the acceleration process. 
The electron phase-space and SPW magnetic field at the target surface ($x_f=25\lambda_0$) are shown at two different times, for (a,b)~$\Delta\beta=0$ and (c,d)~$\Delta\beta=67{\rm mrad}$ (optimal condition). 
In both cases the duration of the electron bunch is proportional to the duration of the SPW, the shortest SPW obtained for $\Delta\beta=67{\rm mrad}$ leading to the shortest electron bunch.
For $\Delta\beta=0$ [panels~(a) and (b)], the SPW is strongly damped at $t=t_0+20\lambda_0/c$: the electron bunch has reached its parallel momentum  ${\rm max}(p_y)\simeq 90 m_e c$ and has a width (measured from the FWHM in momentum) $\Delta y_{fi}=11\lambda_0$. 
The acceleration process is more efficient using the optimal WFR parameter $\Delta\beta=67{\rm mrad}$ [panels (c) and (d)]. At $t=t_0 + 10\lambda_0/c$, two periods after the laser has left the surface, the magnetic field is intense ($\hBspw \simeq 1.2\,a_0$) and the most energetic electrons have already reached momentum up to ${\rm max}(p_y) \simeq 70m_e c$. Ten periods later, a narrow ($\Delta y_{fi}=3\lambda_0$) and energetic [${\rm max}(p_y) \simeq 130m_e c$] electron bunch is obtained,
while the SPW has been significantly damped.

\begin{figure}[t!]
\begin{center}
\includegraphics[width=3.2cm]{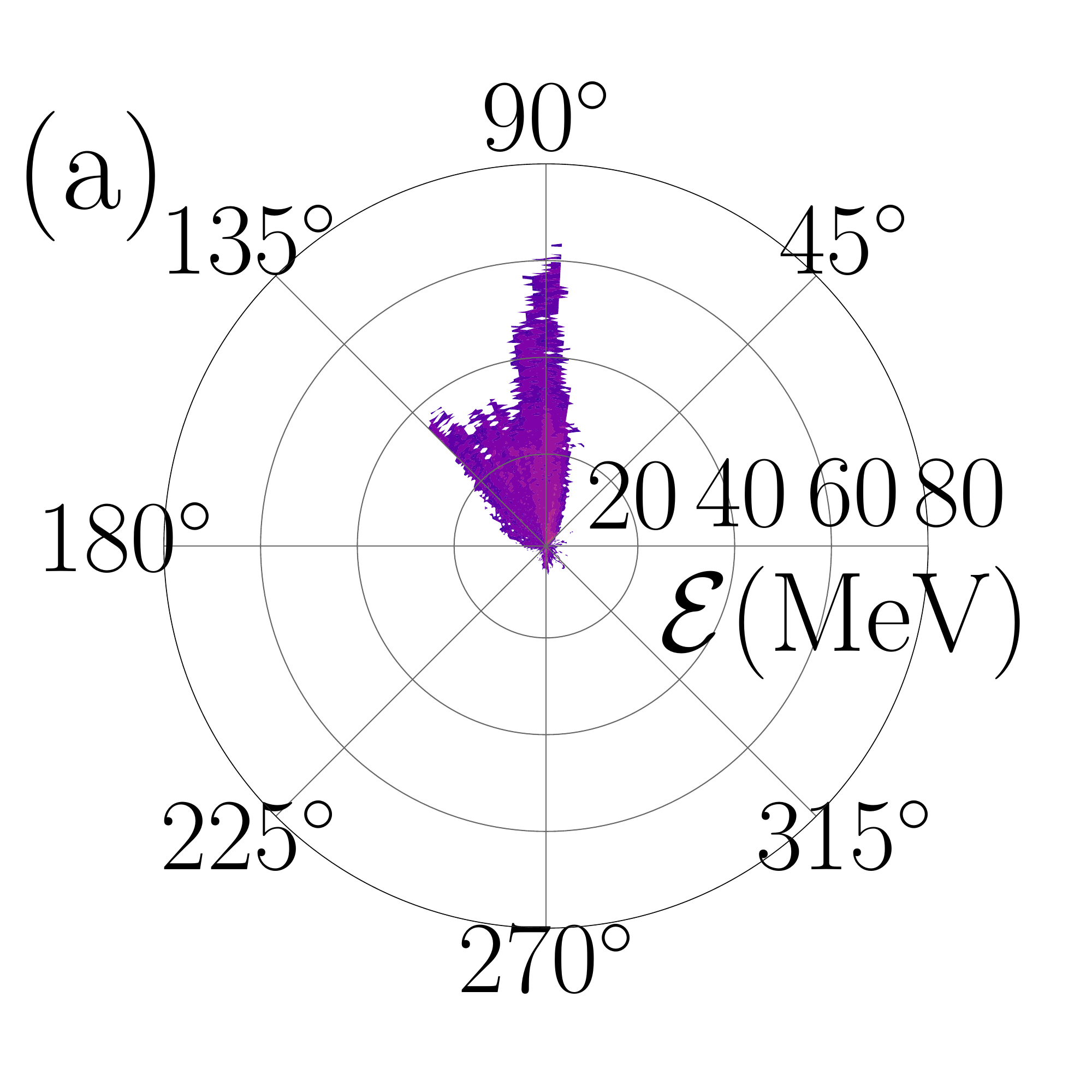}
\includegraphics[width=3.83226cm]{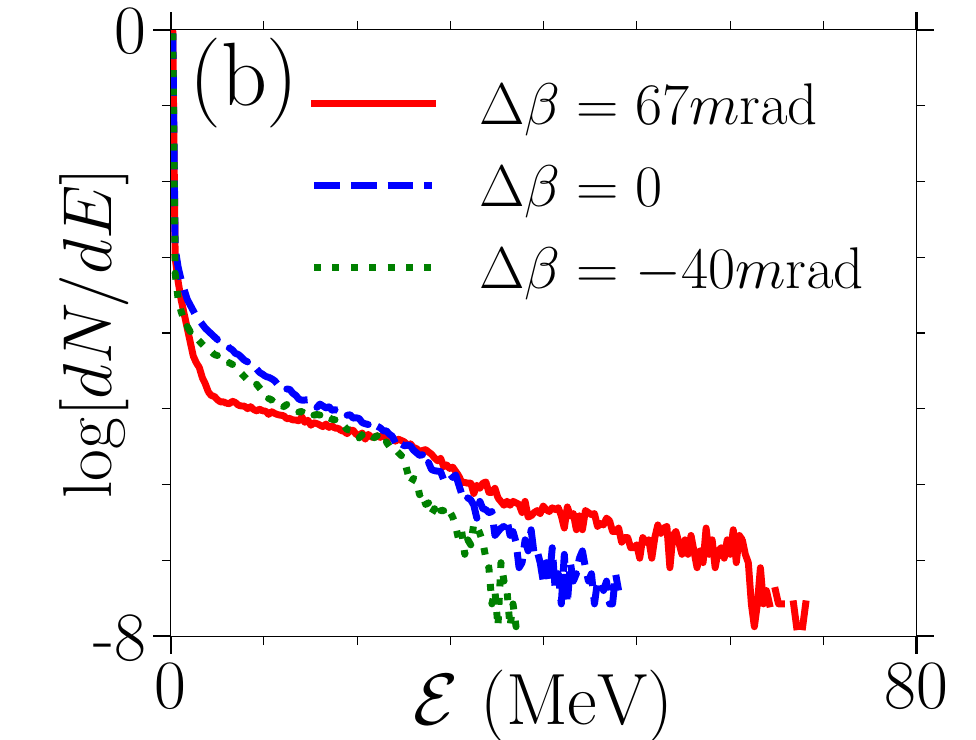}
\caption{(a) Electron energy distribution in MeV as a function of the emission angle $\phi=\tan^{-1}(p_y/p_x)$ for $\Delta\beta=67{\rm mrad}$ and $a_0=5$. (b) Electron energy distribution for $\Delta\beta=67{\rm mrad}$ (red), $\Delta\beta=0$ (blue) and $\Delta\beta=-40{\rm mrad}$ (green).} \label{fig4}
\end{center}
\end{figure}

Similar observations can be drawn from Fig.~\ref{fig4}. In panel (a), the electron distribution in energy and direction (the angle is defined in the simulation plane with respect to the $x$-axis) is shown, 
demonstrating that the most energetic electrons are accelerated mainly along the target's surface and in the $y>0$-direction ({\it i.e.} in the SPW direction of propagation). Panel (b) shows the energy distribution of the electron, for different values of $\Delta\beta$.
%note that the most energetic electrons are propagating along the target surface for $x_f=25\lambda_0$ and considering three values of $\Delta\beta$, with $\Delta\beta = 67{\rm mrad}$ being the optimal case.

These results and in particular the increase of the maximum electron energy (equiv. momentum) 
are consistent with what one expects from the increase of the SPW amplitude by use of the WFR driving pulse. Indeed, an upper limit of the electron energy gain in the SPW has been derived in \cite{riconda} by generalizing the results of wakefield acceleration \cite{tajima,mora}, leading $\Delta {\cal E} \sim \chi\, \gamma_{\phi} \max|\hBspw| m_e c^2$ so that $\Delta {\cal E}$ is proportional to the SPW field amplitude. Here $\gamma_\phi= (1-v_{\phi}^2/c^2)^{-1/2}$ and $\chi$ is a constant of order one, reaching at most 4\cite{riconda}. 
In our simulations, the magnetic field of the SPW (time-averaged over the wave period) reached 
at most $\max|\hBspw| \sim 3.8$ for $\Delta\beta = 0$ and $\max|\hBspw| \sim 7.0$ for $\Delta\beta=67{\rm mrad}$. Considering that $\gamma_\phi\simeq 10$ for $n=100n_c$, 
we then obtain the upper limit $\Delta {\cal E} \simeq 154 m_e c^2$ for $\Delta\beta=0$
and  $\Delta\beta=67{\rm mrad}$ for $\Delta {\cal E} \simeq 280 m_e c^2$. These predictions overestimate the electron energy as they assume i) no wave decay over
the distance required for acceleration, ii) optimal electron injection and iii) acceleration exactly parallel to the target surface, while it has been observed that electrons are deflected in the perpendicular direction \cite{fedeli:16}.

To gain further insight into the acceleration process, we performed a particle tracking of the most energetic electrons and evaluated the trajectory-averaged value of the longitudinal field $\left\langle E_y\right\rangle$ acting on the particle. This allows to define an acceleration length $l_{\rm acc}=\Delta {\cal E}/\left\vert e \left\langle E_y \right\rangle \right\vert$.
From the particle track we found $\Delta {\cal E} \simeq 90 m_e c^2$ and $\left\langle E_y \right\rangle\simeq -1.0\, m_ec\omega_0/e$ for $\Delta\beta=0$, and $\Delta {\cal E} \simeq 130 m_e c^2$ and $\left\langle E_y \right\rangle\simeq -1.4\, m_ec\omega_0/e$ for $\Delta\beta=67{\rm mrad}$. In both cases this leads to an acceleration length $l_{acc} \sim 15\lambda_0$, consistent with the observed particle trajectories. This length largely exceeds the laser spot size, and is close to the length over which the SPW decreases its amplitude significantly (see, {\it e.g.}, Fig.~\ref{fig3}). This confirms the electrons are accelerated by the SPW as it propagates along the target surface.

In the optimal case, the highest energy particles (in the range $30-70 \rm{MeV}$) form a bunch
with duration of $\simeq 3\lambda_0/c$ [$\sim\!\!\!8\,{\rm fs}$ for $\lambda_0=0.8 {\rm \mu m}$] and total charge $\simeq 10 {\rm pC}/\lambda_0$ (in our 2D simulations). Assuming a bunch width (in the $z$-direction) of the order of the laser pulse with $w_{\perp} = 5.2\lambda_0$, one could expect few cycles electron bunches with a charge of $\sim 52 {\rm pC}$. 
These results are competitive with cutting-edge laser wakefield electron beams from underdense plasmas. Considering similar laser parameters and electron energies,  short, high-charge electron bunches were obtained with energy $85$MeV ($21$MeV energy spread), total charge $15$pC and duration $4.4$fs \cite{lundh11}.

In conclusion,  a laser with WFR and an appropriately tailored plasma target  allow to control the duration and amplitude of SPW in the  linear and relativistic regime. 
As a consequence, ultrashort (near single cycle), energetic and highly charged electron bunches are generated. The optimal parameters are clearly identified; since they are well within the capabilities of current UHI installations, this work opens new prospects and provide guidelines for forthcoming experiments.

\vspace{1mm}
Financial support from Grant No. ANR-11-IDEX-0004-02 Plas@Par is acknowledged.
Simulations were performed on the Irene-SKL machine hosted at TGCC- France, using High Performance Computing resources from GENCI-TGCC (Grant No. 2018-x2016057678) and PRACE (project MIMOSAS).
%%TC:endignore

\end{document}